\documentstyle[11pt]{article}
\title{ NEXT-to-LEADING ORDER CONSTITUENT QUARK STRUCTURE AND HADRONIC
STRUCTURE FUNCTIONS
}

\author{Firooz Arash$^{(a,b,c)}$\footnote{farash@cic.aut.ac.ir} and
Ali N. Khorramian$^{(c,d)}$\footnote{khorramiana@theory.ipm.ac.ir}\\
$^{(a)}$
Center for Theoretical Physics and Mathematics, AEOI,\\
P.O. Box 11365-8486, Tehran, Iran \\
$^{(b)}$
Physics Department, AmirKabir University (Tafresh Campus), \\
Hafez Ave.,Tehran Iran 19834 \\
$^{(c)}$
Institute for Studies in Theoretical Physics and Mathematics \\
P.O.Box 19395-5531, Tehran, Iran \\
$^{(d)}$
Physics Department, Semnan University, Semnan, Iran, \\
}
\date{\today}
\begin{document}
\maketitle
\begin{abstract}
We calculate the partonic structure of a constituent quark in the Next-to-Leading
Order framework. The structure of any hadron can be obtained thereafter
using a convolution method. Such a procedure is used to generate the structure
function of proton and pion in NLO, neglecting certain corrections to
$\Lambda_{QCD}$. It is shown that while the constituent
quark structure is generated purely perturbatively and accounts for the most part of the
hadronic structure, there is a few percent contributions coming from the
nonperturbative sector in the hadronic structure. This contribution plays the
key role in explaining the $SU(2)$ symmetry breaking of the nucleon sea and the
observed violation of Gottfried sum rule. These effects are calculated. We obtained
an Excellent agreement with the experimental data in a wide range of $x=[10^{-6}, 1]$ and $Q^{2}=[0.5, 5000]$
$GeV^{2}$ for the proton structure function. We have also calculated
Pion structure and compared it with the existing data. Again, the model calculations
agree rather well with the data from experiment.
\end{abstract}
\section{INTRODUCTION}
Our knowledge of hadronic structure is based on the hadron spectroscopy and
the Deep Inelastic Scattering (DIS) data. In the former picture quarks are massive
particles and their bound states describe the static properties of hadrons.
On the other hand, the interpretation of DIS data relies upon the quarks of QCD Lagrangian
with very small mass. The hadronic structure in this picture is intimately
connected with the presence of a large number of partons (quarks and gluons).
Mass is not the only difference between these
two types of quarks; they also differ in other important
properties. For example, the color charge of quark field in QCD Lagrangian is
ill-defined and is not gauge invariant, reflecting the color of gluons in an
interacting theory; whereas, color associated with a constituent
quark (CQ) is a well defined entity. It is shown that \cite{1}\cite{2}
one can perturbatively dress a QCD Lagrangian field to all orders and construct
a CQ in conformity with the color confinement. From this point of view a CQ is
defined as a quasi-particle emerging from the dressing of valence quark with
gluons and $q-\bar{q}$ pairs in QCD.\\
Of course, the concept of CQ as an intermediate step between the
quarks of QCD Lagrangian and hadrons is not new. In fact,
Altarelli and Cabibo \cite{3} used this concept in the context of
$SU(6) X O(3)$. R.C. Hwa in his work used the term
{\it{valon}},for them, elaborated on the concept and showed its
applications to many physical processes \cite{4}. In Ref.[2] it is
suggested that the concept of dressed quark and gluon might be
useful in the area of jet physics and heavy quark effective
theory. Despite the ever presence of CQ no one has calculated its
content and partonic structure without resorting to hadronic data
and the process of deconvolution. Thus, the purpose of this paper
is threefold: (a) to evaluate the structure of a CQ in the
Next-to-Leading Order (NLO) framework of QCD, (b) to verify its
conformity with the structure function data of nucleon and pion
for which ample data are available, (c) to include additional
refinments that are needed to account for the violation of
Gottfried Sum Rule (GSR) and the
binding effect of constituent quarks to form a physical hadron. \\
\section{FORMALISM}
\subsection{Moments of Parton Distribution in a CQ}
In this subsection we will utilize the extended work done on the development
of the NLO calculations of moments, to calculate the structure of a
constituent quark.
What will follow in the rest of this subsection is not a new next-to-leading
order calculation. However, we find it interesting to explore the existing
calculations in the valon framework.\\
By definition, a CQ is a universal building block for every hadron; that is,
the structure of a CQ is common to all hadrons and is generated perturbatively.
Once its structure is calculated, it would be possible, in principle, to
calculate the structure of any hadron. In doing so, we will follow the philosophy
that in a DIS experiment
at high enough $Q^{2}$, it is the structure of a CQ which is being probed,
and that at
sufficiently low value of $Q^{2}$ this structure cannot be resolved. Thus, a CQ
behaves as a valence quark, and a hadron is viewed as the bound state of its
constituent quarks. Under these criteria, partons of DIS experiments are
components of CQ. The structure function of a U-type CQ at high $Q^{2}$ can
be written as[4]:
\begin{equation}
F_{2}^{U}(z,Q^2)=\frac{4}{9}z(G_{\frac{u}{U}}+G_{\frac{\bar{u}}{U}})+ \frac{1}
{9}z(G_{\frac{d}{U}}+G_{\frac{\bar{d}}{U}}+G_{\frac{s}{U}}+G_{\frac{\bar{s}}{U}})+...
\end{equation}
where all the functions on the right-hand side are the probability functions for
quarks having momentum fraction $z$ of a U-type CQ at $Q^{2}$. A
similar expression can be written for a D-type CQ.
Following Ref. [4], we define the singlet (S) and nonsinglet (NS) components of
the CQ structure functions as:
\begin{equation}
G^{S}=\sum^{f}_{i=1}(G_{\frac{q_{i}}{CQ}} + G_{\frac{\bar{q}_{i}}{CQ}}) =G_{f} +(2f-1)G_{uf}
\end{equation}
\begin{equation}
G^{NS}=\sum^{f}_{i=1}(G_{\frac{q_{i}}{CQ}} -G_{\frac{\bar{q}_{i}}{CQ}}) =G_{f} - G_{uf}
\end{equation}
where $G_{f}$ is the favored distribution describing the structure function of
a quark within a CQ of the same flavor, while the unfavored distribution, $G_{uf}$,
describes the structure function of any quark of different flavor within the CQ.
We have used, $f$, for the number of active flavors. Solving (2) and (3) for
$G_{f}$ and $G_{uf}$, we get,
\begin{equation}
G_{f}=\frac{1}{2f}(G^{S} + (2f-1) G^{NS}),
\end{equation}
\begin{equation}
G_{uf}=\frac{1}{2f}(G^{S}-G^{NS}).
\end{equation}
Having expressed all the structure functions of a CQ in terms of $G_{f}$ and
$G_{uf}$, we now go to the N- moment space and define the moments of these
distributions as:
\begin{equation}
M_{2}(N,Q^{2})=\int_{0}^{1} x^{N-2} F_{2}(x,Q^{2}) dx
\end{equation}
\begin{equation}
M_{i}(N,Q^{2})=\int_{0}^{1} x^{N-1} G_{i}(x,Q^{2})dx
\end{equation}
where, the subscription $i$ stands for S or NS. Due to charge symmetry, in
the following, we only write the CQ distribution for proton. \\
In the NLO approximation the dependence of the running coupling constant,
$\alpha_{s}$, on $Q^{2}$ is given by:
\begin{equation}
\alpha_{s}(Q^{2})\approx \frac{4\pi}{\beta_{0}ln(\frac{Q^{2}}{\Lambda^{2}})}
\big(1-\frac{\beta_{1}ln ln(\frac{Q^{2}}{\Lambda^{2}})}
{\beta_{0}^{2}ln(\frac{Q^{2}}{\Lambda^{2}})}\big),
\end{equation}
with $\beta_{0}=\frac{1}{3} (33-2f)$ and $\beta_{1}=102-\frac{38f}{3}$. The
evolution of parton distributions is controlled by the anomalous dimensions: \\
\hspace{2cm} $\gamma^{N}_{NS}=\frac{\alpha_{s}}{4\pi}\gamma^{(0)N}_{qq}+ (\frac{\alpha_{s}}{4\pi})^{2}\gamma^{(1)N}_{NS}$
\hspace{3cm}
$\gamma^{N}_{ij}=\frac{\alpha_{s}}{4\pi}\gamma^{(0)N}_{ij}+ (\frac{\alpha_{s}}{4\pi})^{2}\gamma^{(1)N}_{ij}$ \\
where $i, j =q, g$; and $\gamma^{(0,1)N}$ are one and two loop
anomalous dimension matrices. The nonsinglet moments in the NLO
are as follows:
\begin{equation}
M^{NS}(N,Q^{2})= [1+\frac{\alpha_{s}(Q^{2})-\alpha_{s}(Q_{0}^{2})}{4\pi}\big(
\frac{\gamma^{(1)N}_{NS}}{2\beta_{0}}-\frac{\beta_{1}\gamma_{qq}^{(0)N}}{2\beta_{0}^{2}}
\big)](\frac{\alpha_{s}(Q^{2})}{\alpha_{s}(Q_{0}^{2})})^{
\frac{\gamma_{qq}^{(0)N}}{2\beta_{0}}}.
\end{equation}
The anomalous dimension matrices, $\gamma^{(0,1)N}$,are given by,
\begin{equation}
\gamma^{(0,1)N}=\left( \begin{array}{cc} \gamma^{(0,1)N}_{qq} &
\gamma^{(0,1)N}_{qg} \\ \gamma^{(0,1)N}_{gq} &
\gamma^{(0,1)N}_{gg}
\end{array} \right)
\end{equation}
These matrices, govern the moments of the singlet quark and sector as given by:
\begin{eqnarray}
 \left( \begin{array}{c}
 M^{S}(N,Q^{2}) \\ M^{G}(N,Q^{2})
 \end{array} \right)
=\{(\frac{\alpha_{s}(Q^{2})}{\alpha_{s}(Q_{0}^{2})})^{\frac{\lambda_{-}^{N}}{2\beta_{0}}}[
p_{-}^{N}-\frac{1}{2\beta_{0}}\frac{\alpha_{s}(Q^{2}_{0})-\alpha_{s}(Q^{2})}{4\pi}p_{-}^{N}
\gamma^{N}p_{-}^{N}-\nonumber\\
\big(\frac{\alpha_{s}(Q^{2}_{0})}{4\pi}-\frac{\alpha_{s}(Q^{2})}{4\pi}(\frac{\alpha_{s}(Q^{2})}{\alpha_{s}(Q_{0}^{2})}
)^{\frac{\lambda_{+}^{N}-\lambda_{-}^{N}}{2\beta_{0}}}\big)
\frac{p_{-}^{N}\gamma^{N}p_{+}^{N}}{2\beta_{0}+\lambda^{N}_{+}-\lambda^{N}_{-}}]\}{\bf{1}}+\{({\cal{Q}}_{+\longleftrightarrow
-})\}
\end{eqnarray}
where
$\gamma^{N}=\gamma^{(1)N}-\frac{\beta_{1}}{\beta_{0}}\gamma^{(0)N}$
and $\gamma^{(0)N}$, $\gamma^{(1)N}$ are one and two-loop
anomalous dimension matrices, respectively. ${\bf{1}}$ is the unit
matrix. Notice that the last term in Eq. (11),
${\cal{Q}}_{+\longleftrightarrow -}$, means that all subscripts
are exchanged. The leading order behavior is obtained from the
first term in the square brackets. $\lambda_{\pm}^{N}$ denote the
eigenvalues of the one-loop anomalous dimension matrix,
$\gamma^{(0)N}$:
\begin{equation}
\lambda_{\pm}^{N}=\frac{1}{2}[\gamma_{qq}^{(0)N}+\gamma_{gg}^{(0)N}\pm \sqrt{
(\gamma_{gg}^{(0)N}-\gamma_{qq}^{(0)N})^{2}+4\gamma_{qg}^{(0)N}\gamma_{gq}^{(0)N}}]
\end{equation}
These anomalous dimensions and $\gamma_{kl}^{(1,0)N}$ are calculated
up to second order in $\alpha_{s}$ and can be found in \cite{5} \cite{6}.
$p_{\pm}^{N}$ are as follows:
\begin{equation}
p_{\pm}^{N}=\pm(\gamma^{(0)N}-\lambda_{\mp}^{N})/(\lambda_{+}^{N}-\lambda_{-}^{N})
\end{equation}
In Ref. [4], $d_{\pm}^{(0)}$ is used to denote the leading order
anomalous dimensions; these are related to $\lambda_{\pm}$ by:
$d_{\pm}^{0}=\frac{\lambda_{\pm}^{N}}{2\beta_{0}}$. \\
In evaluating these moments, we have taken $Q^{2}_{0}=0.283$ $GeV^{2}$ and $\Lambda=0.22$ $GeV$,
as our initial scales. It seems that evolution of parton distributions
from such a low value of $Q^{2}_{0}$ is not justified theoretically;
we shall take up this issue shortly in subsection 2.3.
\subsection{Parton Densities In CQ}
In the previous subsection we outlined how to calculate the moments. In this
subsection we will give the corresponding parton distributions in a CQ.\\
The moments of the CQ structure function, $F_{2}^{CQ}(z,Q^{2})$, are expressed
completely in terms of $Q^{2}$ through the evolution parameter $t$:
\begin{equation}
t={\it{ln}}\frac{\it{ln}\frac{Q^2}{\Lambda^2}}{\it{ln}\frac{Q_{0}^{2}}{\Lambda^2}}
\end{equation}
The moments of valence and sea quarks in a CQ are,
\begin{equation}
M_{\frac{valence}{CQ}}=M^{NS}(N,Q^{2})
\end{equation}
\begin{equation}
M_{\frac{sea}{CQ}}=\frac{1}{2f}(M^{S}-M^{NS})
\end{equation}
where $M^{S, NS}$are given by eqs. (9) and (11). It is now straightforward,
to evaluate $M_{\frac{valence}{CQ}}$ and $M_{\frac{sea}{CQ}}$ at any $Q^{2}$
or $t$. These are calculated numerically for a range of $N$ at fixed $Q^{2}$ or $t$.
In Figure (1) we present these moments both in the leading and next-to-leading
orders. For every value of $t$ we fit the moments by a beta function, which in
turn is the moment of the parton distribution as given below,
\begin{equation}
zq_{\frac{val.}{CQ}}(z,Q^{2})=a z^{b}(1-z)^{c}
\end{equation}
\begin{equation}
zq_{\frac{sea}{CQ}}(z,Q^{2}) = \alpha z^{\beta}(1-z)^{\gamma}[1+\eta z +\xi z^{0.5}]
\end{equation}
The parameters $a$,$b$, $c$, $\alpha$, $\beta$, $\gamma$, $\eta$,
and $\xi$ are functions of $Q^{2}$ through the evolution parameter
$t$. The same form as in Eq. (18) is obtained for the gluon
distribution in a CQ, only with different parameter values. The
functional forms of these parameters are polynomials of order
three in $t$ and are given in the appendix. The goodness of these
we fits are checked by $\chi^{2}$ minimization procedure. We find
that, for all parameters $a$, $b$, $c$, and $d$, of the valence
quark sector $\chi^{2}/DOF=0.99$, with the standard error of order
$10^{-3}$. For the sea quark and gluon sector,  $\chi^{2}/DOF$
carries between $0.91$ and $0.98$. Of course, the forms (17) and
(18) are not unique; rather, they are the most simple and commonly
used forms. We note that the sum rule reflecting the fact that
each CQ contains only one valence quark is satisfied at all values
of $Q^{2}$:
\begin{equation}
\int^{1}_{0} q_{\frac{val.}{CQ}}(z,Q^{2}) dz = 1.
\end{equation}
Substituting these results into Eq. (1) completes the evaluation of CQ
structure function in NLO. In Figure (2), we plot various parton
distributions inside a CQ.
\subsection{discussion}
In subsection 2.1 we indicated that our initial scales are $Q_{0}^{2}=0.283$ $GeV^{2}$
and $\Lambda=0.22$ $GeV$. The above value of $Q_{0}$ corresponds to a distance
of 0.36 {\it{fm}} which is
roughly equal to or slightly less than the radius of a CQ. It may be objected
that such distances are probably too large for a meaningful pure perturbative
treatment. We note that $F_{2}^{CQ}(z,Q^{2})$ has the property that it becomes
$\delta(z-1)$ as $Q^{2}$ is extrapolated to $Q_{0}^{2}$, which is beyond the
region of validity. This mathematical boundary condition signifies that the
internal structure of a CQ cannot be resolved at $Q_{0}$ in the NLO approximation.
Consequently, when this property is applied to Eq.(20), below, the structure function
of the nucleon becomes directly related to $xG_{\frac{CQ}{P}}(x)$ at those
values of $Q_{0}$; that is, $Q_{0}$ is the leading order effective value at
which the hadron can be considered as consisting of only three (two) CQ, for
baryons (mesons). We have checked that when $Q^{2}$
approaches to $Q_{0}^{2}$ the quark moments approach to unity and
the moments of gluon approach to zero. In fact for $Q^{2}=0.2839$
gluon moments are at the order of $10^{-4}$ and singlet and
nonsinglet moments are 0.9992. In fact our results are only meaningful
for $Q_{0}^{2}\ge 0.4$ $GeV^{2}$. \\
>From the theoretical standpoint, both $\Lambda$ and $Q_{0}$ depend on the
order of the moments, $N$; but here, we have assumed that they are
independent of $N$. In this way, we have introduced some degree of
approximation to the $Q^{2}$ evolution of the valence and sea quarks. However,
on one hand there are other contributions like target-mass effects, which add
uncertainties to the theoretical predictions of perturbative QCD, while on the
other hand since we are dealing with the CQ, there is no experimental data
to invalidate an $N$ independent $\Lambda$ assumption.
\section{HADRONIC STRUCTURE}
\subsection{Constituent Quark Distribution in hadron}
So far, the structure of a CQ is calculated in the NLO framework.
In this section we will use the convolution method to calculate the structure
functions of proton, $F_{2}^{p}(x,Q^{2})$, and pion. Let us denote the structure
function of a CQ by $F^{CQ}_{2}(z,Q^{2})$ and the probability of finding a CQ
carrying momentum fraction $y$ of the hadron by $G_{\frac{CQ}{h}}(y)$. The
corresponding structure function of a hadron is obtained by convolution of
$F^{CQ}_{2}(z,Q^{2})$ and $G_{\frac{CQ}{h}}(y)$,
\begin{equation}
F_{2}^{h}(x,Q^2)=\sum_{CQ}\int_{x}^{1}dy G_{\frac{CQ}{h}}(y)F^{CQ}_{2}(\frac{x}{y},Q^2)
\end{equation}
where summation runs over the number of CQ's in a particular hadron. We note
that $G_{\frac{CQ}{h}}(y)$ is independent of the nature of the probe, and its
$Q^{2}$ value. In effect $G_{\frac{CQ}{h}}(y)$ describes the wave function of
hadron in CQ representation, containing all the complications due to confinement.
>From the theoretical point of view, this function cannot be evaluated accurately.
To facilitate phenomenological analysis, following [4], we assume a simple
form for the exclusive CQ distribution in proton and pion as follows,
\begin{equation}
G_{UUD/p}(y_{1},y_{2},y_{3})=l(y_{1}y_{2})^{m}y_{3}^{n}\delta(y_{1}+y_{2}+y_{3}-1)
\end{equation}
\begin{equation}
G_{\bar{U}D/\pi^{-}}(y_{1},y_{2})=qy_{1}^{\mu}y_{2}^{\nu}\delta(y_{1}+y_{2}-1)
\end{equation}
where $l$ and $q$ are normalization parameters. After integrating over unwanted
momenta, we can arrive at the inclusive distribution of individual CQ:
\begin{equation}
G_{U/p}(y)=\frac{1}{B(m+1,n+m+2)}y^{m}(1-y)^(m+n+1),
\end{equation}
\begin{equation}
G_{D/p}(y)=\frac{1}{B(n+1,2m+2)}y^{n}(1-y)^{2m+1},
\end{equation}
\begin{equation}
G_{\bar{U}/\pi^{-}}(y)=\frac{1}{B(\mu +1, \nu +1)}y^{\mu}(1-y)^{\nu}.
\end{equation}
Similar expression for $G_{D/\pi^{-}}$ is obtained with the interchange of
$\mu\leftrightarrow \nu$. In the above equations, $B(i,j)$ is the Euler Beta function.
The arguments of this function, as well as, $l$ and $q$ of Eqs.(21) and (22),
are fixed by the number and momentum sum rules:
\begin{equation}
\int^{1}_{0}G_{\frac{CQ}{h}}(y) dy=1 , \hspace{2cm}\sum_{CQ}\int^{1}_{0}yG_{\frac{CQ}{h}}(y) dy=1 ,
\end{equation}
where $CQ=U, D, \bar{U}$ and $h=p, \pi^{-}$. Numerical values are:
$\mu=0.01$, $\nu =0.06$, $m=0.65$ and $n=0.35$. These parameters are independent
of $Q^2$ and are given in [4] and \cite{7} for proton and pion, respectively.
In \cite{8}, a new set of values for $m$ and $n$ are suggested which differ
significantly from the values quoted above. Reference [8] uses the CTEQ parton
distributions in the Next-to-Leading order to fix these parameters.
The authors of Ref. [8] have found yet another set of values in
\cite{9}, which differs from those given in [4] and [8]. They attribute
these differences to the different theoretical assumptions. We
have tried a range of values for $m$, $n$, $\nu$ and $\mu$, and checked
that the sum rules (26) are satisfied.
Because of the Beta function in the denominator of Eqs.(23-25) the
number sum rule does not change by changing $m$ and $n$ and we
found little sensitivity on the momentum sum rule. In Fact, we have
varied both $m$ and $n$, randomly in the interval $[0.2, 1.95]$, and
checked it against $F_{2}^{p}$ data. We found very little sensitivity to the
choices made. For the data that we shall analyze, these two parameter formulas
will prove to be adequate. The differences between parameters of Ref. [8] and
Ref. [4] (which we have adopted) can be attributed to the different methods
of extracting the parameters. We have calculated the parton distributions
from the perturbative QCD, whereas Ref.[8] have used CTEQ's global fit to
extract parton content of nucleon.  \\
The considerations on the form of CQ distribution that are used here,
does not exclude other possibilities. However, in our model the quark
distribution, say, in a proton is described as the convolution of the CQ
distribution, $G_{\frac{CQ}{p}}$, and the quark distribution in the CQ. In the
moment form the latter two become the product of their moments. Mathematically,
these two moments can be modified without changing the product. That would,
of course, modify the original $G_{\frac{CQ}{p}}$ and parton distribution in a
CQ. But the parton distribution has a definite meaning. It is the evolution of
a quark as $Q$ increases. At $Q=Q_{0}$, parton distribution is a delta function,
implying that the CQ has no internal structure that can be probed. Our model is
to say that the three valence quark at high $Q$ becomes three CQ's at low $Q$
without anything else. Such a physical statement of the model does not permit
to redefine $G_{\frac{CQ}{p}}$ or the parton distribution in a CQ, although
mathematical criticism still remains. \\
Our choice of $G_{\frac{CQ}{p}}$ is sufficient to fit all the data
of deep inelastic scattering in the range of interest. It does not
mean that a more complicated $G_{\frac{CQ}{p}}$ is ruled out. It
is a subject that can be revisited in the future when problem
arise. At this point our formula for the CQ distribution is
satisfactory. A comment regarding the moments is in order: The
definition of the moments are not arbitrary if we are to attach
physical interpretation to the constituent quark distribution.
There are only two U and one D in a proton, which impose
constrains on the first moment. Also, the sum of the CQ momenta
should be one, which imposes a constraint on the second moment. If
one writes the valence quark moments as a product of $G_{n}$ and
$M^{NS}$ then $M^{NS}$ must have connection with QCD evolution to
ensure that there is only one valence quark in every CQ and hence,
the constraint becomes $M_{1}^{NS}=1.$ With these constrains, one
cannot freely redefine moments of CQ distribution, $M^{NS}$ and
$M^{S}$. yet, there are infinite number of CQ moments and we have
cited only two constraints. Does this leave an infinite horde
number of undefined quantities? The higher moments of the CQ
distribution are affected mainly by the large $x$ behavior of
proton wave function, or will affect only the even larger $x$
behavior of the parton distribution. Until one focuses on that
region and finds inconsistency, one chooses the simplest
parameterization. There are similar situations both in classical
and contemporary physics. The question can be turned around and
asked why did Newton stop at $\frac{1}{r^{2}}$ law for
gravitational force? Could he have excluded the possible existence
of $\frac{1}{r^{n}}$ terms with $n= 3,4,5,...$? By only citing the
existence of two constraints, Newton avoided confronting the
infinite horde of undefined quantities. Our model is inductive
physics. Valons or CQ's are not derived by deductive logic from
the first principles. In modeling, one finds the simplest way to
capture the essence of the physics involved. The situation is also
present in contemporary physics: we could have asked why the
Cornell potential for the bound-state problem of quark and
antiquark does not contain all possible power law terms beside the
$\frac{1}{r}$ and the $r$ terms. The answer is simple: it is what
fits the particle mass spectrum that counts, again
in the simplest way possible that captures the essence of the problem.\\
\subsection{Parton distribution in hadrons}
Having specified the parton distributions in the CQ and the constituent quark
distribution in proton and pion, it is possible to determine various parton
distributions in any hadron. For proton, we get:
\begin{eqnarray}
\begin{array}{c}
q_{val./p}(x,Q^{2})=\nonumber \\
\end{array}
2\int^{1}_{x}\frac{dy}{y} G_{U/p}(y) q_{val./U}(\frac{x}{y},Q^{2}) \nonumber \\
+\int^{1}_{x}\frac{dy}{y} G_{D/p}(y) q_{val./D}(\frac{x}{y},Q^{2}) =u_{val./p}(x, Q^{2}) +d_{val./p}(x, Q^{2})
\end{eqnarray}
\begin{equation}
q_{sea/p}(x,Q^{2})=2\int^{1}_{x}\frac{dy}{y} G_{U/p}(y) q_{sea/U}(\frac{x}{y},Q^{2}) +
\int^{1}_{x}\frac{dy}{y} G_{D/p}(y) q_{sea/D}(\frac{x}{y},Q^{2})
\end{equation}
The above equations represent the contribution of constituent
quarks to the sea and the valence quark distributions in proton.
In Figure (3), the CQ contribution to parton distributions in $p$
and $\pi^{-}$ is shown. Comparing with proton structure function
data, shows that the results fall short of representing the
experimental data by a few percent ($3\%$-$8\%$, depending on
$Q^{2}$ and $x$). This is not surprising, for it is a well
established fact that there are soft gluons in the nucleon. We
attribute their presence to the fact that constituent quarks are
not free in a hadron, but interact with each other to form the
bound states. At each $Q^{2}$ value we have computed the soft
gluon contribution using direct comparison with experimental data.
The final result is parameterized as:
\begin{equation}
xg_{soft}=R(Q^{2})x^{-0.029}(1-x)^{2.79};
\end{equation}
where, $R(Q^{2})=2.14-6.13t+6.177t^{2}-1.812t^{3}$. In figure (4), the shape of
$xg_{soft}$ is presented for a range of $Q^{2}$ values. Splitting of these
gluons could in turn produce $\bar{q}-{q}$ pairs which can combine with the
original constituent quarks and fluctuate the baryonic state to a meson-baryon
state. This also will break the $SU(2)$ symmetry of the sea as we shall see next.\\
\subsection{The Role of Soft Gluon}
In our model there is no room in the CQ structure for the breaking
of $SU(2)$ symmetry of the nucleon sea. But, after creation of
$\bar{q}-q$ pair by the soft gluons, these quarks can recombine
with CQ to fluctuate into meson-nucleon state which breaks the
symmetry of the nucleon sea. In what follows we will compute this
component, and its contribution to the nucleon structure function,
and its relevance to the violation of the Gottfried sum rule. We
will follow the prescription of \cite{10} and \cite{11}. The above
proposed model, is similar to the effective chiral quark theory,
in which the relevant degrees of freedom are constituent quarks,
gluons and Goldstone bosons. The coupling of pions produces pion
cloud of the constituent quark, in which we are interested. The
point we would like to make is that, there are two types of sea
quark and gluon that contribute to the nucleon sea measured in
DIS. One type is generated from QCD hard bremsstrahlung and gluon
splitting. This component is discussed in subsection 2.2 and is
associated with the internal composition of the constituent quark
rather than the proton itself. The other contribution is generated
nonperturbatively via meson-baryon fluctuation. As we will show,
according to our results, this component provides a consistent
framework to understand the flavor asymmetry of the nucleon sea.
It also compensates for the shortcoming of perturbatively
generated parton densities in reproducing the experimental data on
$F_{2}^{p}$. In order to distinguish these partons from those
confined inside the CQ, we will term them as {\it{inherent}}
partons. This component is intimately related to the bound state
problem, and hence it has a non-perturbative origin. However, for
the process of $CQ\rightarrow CQ+{\it{gluon}}\rightarrow
\bar{q}-q$, at an initial value of $Q^{2}=0.65 GeV^{2}$ where
$\alpha_{s}$ is still small enough, we will calculate it
perturbatively. The corresponding splitting functions are as
follows,
\begin{equation}
P_{gq}(z)=\frac{4}{3}\frac{1+(1-z)^{2}}{z},
\end{equation}
\begin{equation}
P_{qg}(z)=\frac{1}{2}(z^{2}+(1+z)^{2}).
\end{equation}
For the joint probability distribution of the process at hand, we get,
\begin{equation}
q_{inh.}(x,Q^2)=\bar{q}_{inh}(x,Q^2)={\cal{N}}\frac{\alpha_{s}}{(2\pi)}\int_{x}^{1}\frac{dy}{y}P_{qg}(\frac{x}{y})g_{soft}(y)dy
\end{equation}
where $q_{inh.}$ is the inherent parton density.
The splitting functions, and the $q_{inh}(x,Q^2)=(\bar{q}_{inh}(x,Q^2))$,
are those of the leading order, rather than NLO. We do not expect that this
approximation will affect the whole structure function, as can be seen
from Figure (5). In the above equation ${\cal{N}}$, is a factor
depending on $Q^{2}$, and $G_{CQ}$ is the constituent quark distribution
in the proton as given previously. The same process, can also be a source of
$SU(2)$ symmetry breaking of nucleon sea; namely, $u_{sea}\neq d_{sea}$, and
hence, the violation of the Gottfried sum rule (GSR). We will take up this
issue in the next subsection. \\
The process is depicted in Figure (6). Probability to form a meson-baryon state
can be written as in Ref. [11],
\begin{equation}
P_{MB}(x)=\int^{1}_{0}\frac{dy}{y}\int^{1}_{0}\frac{dz}{z}F(y,z)R(y,z;x)
\end{equation}
where $F(y,z)$ is the joint probability of finding a CQ with
momentum fraction $y$, and an inherent quark or anti-quark of
momentum fraction $z$ in the proton. $R(y,z;x)$ is the probability
of recombining a CQ of momentum $y$, with an inherent quark of
momentum $z$, to form a meson of momentum fraction $x$ in the
proton. For a more general case, the evaluation of both of these
probability functions are discussed in \cite{12}. An earlier and
pioneering version was proposed in \cite{13}. In the present case,
these functions are much simpler. Guided by the works of Ref. [11,
12, 13], we can write,
\begin{equation}
F(y,z)=\Omega yG_{\frac{CQ}{p}}(y)z\bar{q}_{inh.}(z)(1-y-z)^{\delta}
\end{equation}
\begin{equation}
R(y,z;x)=\rho y^{a}z^{b}\delta(y+z-1)
\end{equation}
Here, we take $a=b=1$ reflecting that the two CQ's in meson share its momentum
almost equally. The exponent $\delta$ is fixed for the $n$ and $\Delta^{++}$
states, using the data from E866 experiment [14], and the mass ratio of $\Delta$
to $n$. They turn out to be approximately 18 and 13, respectively. $\Omega$
and $\rho$ are the normalization constants, also fixed by data. We recognize
that the discussion leading to Eqs. (34, 35) and fixing the parameters
are based on a phenomenological ansatz. The ansatz and the parameters of Eqs.
(34, 35) seems working remarkably well, and further effects that might be of
hadronic origin need not be larger than a few percent. \\
Now it is possible to evaluate $\bar{u}_{M}$ and $\bar{d}_{M}$ quarks,
associated with the formation of meson states[11]:
\begin{equation}
\bar{d}_{M}(x,Q^{2}) =\int^{1}_{x}\frac{dy}{y}[P_{\pi^{+} n}(y) +
\frac{1}{6}P_{\pi^{+} \Delta ^{0}}(y)]\bar{d}_{\pi}(\frac{x}{y},Q^{2})
\end{equation}
\begin{equation}
\bar{u}_{M}(x,Q^{2})=\frac{1}{2}\int^{1}_{x}\frac{dy}{y}
P_{\pi^{-} \Delta^{++}}(y)\bar{u}_{\pi}(\frac{x}{y},Q^{2})
\end{equation}
where $\bar{u}_{\pi}$ and $\bar{d}_{\pi}$ are the valence quark
densities in the pion at scale $Q^{2}_{0}$. The coefficients $\frac{1}{2}$
and $\frac{1}{6}$ are due to isospin consideration. Using Eqs. (17), (18), and
(25), we can calculate various parton distributions in a pion. Those pertinent to
Eqs. (35) and (36) are:
\begin{equation}
\bar{u}_{val.}^{\pi^{-}}(x, Q^{2})=\int^{1}_{x}G_{\bar{U}/\pi^{-}}(y)
\bar{u}_{val./\bar{U}}(\frac{x}{y},Q^{2})\frac{dy}{y},
\end{equation}
\begin{equation}
d_{val.}^{\pi^{-}}(x, Q^{2})=\int^{1}_{x}G_{D/\pi^{-}}(y)
d_{val./D}(\frac{x}{y},Q^{2})\frac{dy}{y},
\end{equation}
\begin{equation}
\bar{u}_{val./\bar{U}}=u_{val./U}.
\end{equation}
There are some data on the valence structure function of $\pi^{-}$ \cite{15},
which can serve as a check for the proposed model.
Defining valence structure function of $\pi^{-}$ as
\begin{equation}
F^{\pi^{-}}_{val.}=x\bar{u}^{\pi^{-}}_{val.} =xd^{\pi^{-}}_{val.}
\end{equation}
We present the results of our calculation for $F^{\pi^{-}}_{val.}$ in Figure
(7), along with the experimental data at $Q^{2}$ around 6 $Gev^{2}$.
\subsection{Asymmetry of the Nucleon Sea and the Gottfried Sum Rule}
There are experimental evidences \cite{14},\cite{16} that the Gottfried
integral,
\begin{equation}
S_{G}=\int_{0}^{1}[F_{2}^{p}(x)-F_{2}^{n}(x)]\frac{dx}{x}=
\frac{1}{3}-\frac{2}{3}\int_{0}^{1}dx [\bar{d}(x)-\bar{u}(x)],
\end{equation}
is less than $\frac{1}{3}$ which is the value expected in the
simple quark model. The NMC collaboration \cite{16}, result at
$Q^{2}=4$ $GeV^{2}$ is, $S_{G}=0.235 \pm 0.026$, which is
significantly smaller than $\frac{1}{3}$. There are several
explanations for this observed violation of the Gottfreid sum rule
(GSR). Among them are: flavor asymmetry of the nucleon sea, that
is, $\bar{u}_{sea}\neq\bar{d}_{sea}$,\cite{17} \cite{18}, isospin
symmetry breaking between proton and neutron, and Pauli blocking,
among others. One of these
explanations fits well within our model. \\
It was proposed by Eichten, Inchliffe, and Quigg \cite{19}, that valence quark
fluctuates into a quark and a pion. This explanation fits very well in our
model. In other words, a nucleon can fluctuate into a meson-nucleon state. The
idea is appealing to us and in our model it can be calculated rather easily.
According to our model, after a pair of {\it{inherent}} $q-\bar{q}$, is created,
a $\bar{u}$ can couple to a D-type CQ to form
an intermediate $\pi^{-}=D\bar{u}$, while the $u$ quark combines with the other
two U-type CQ's to form a $\Delta^{++}$. This is the lowest $u\bar{u}$ fluctuation.
Similarly, a $d\bar{d}$ can fluctuate into the $\pi^{+}n$ state. Since the
$\Delta^{++}$ state is more massive than the $n$ state, the probability of $d\bar{d}$
fluctuation will dominate over $u\bar{u}$ fluctuation, which naturally leads
to an excess of $d\bar{d}$ pairs over $u\bar{u}$ in the proton sea. This process is
depicted in Figure (6).\\
To summarize, there are three sources that contribute to the sea partons in the
proton: Constituent quarks of the proton, the mesonic cloud of the CQ, and the
splitting of soft gluon. The combined contribution to the ratio of $\frac{\bar{d}}{\bar{u}}$
is as follows:
\begin {equation}
(\frac{\bar{d}}{\bar{u}})_{proton}=\frac{\bar{d}_{M}+\bar{d}_{inh.+CQ}}
{\bar{u}_{M}+\bar{u}_{inh.+CQ}}.
\end{equation}
NUSea collaboration at FermiLab, E866 experiment, has published
their results [14], for the integral of $\bar{d}-\bar{u}$ and
$\frac{\bar{d}}{\bar{u}}$ at $Q=7.35$ $GeV$. With the procedure
described, we have calculated these values at the same $Q$ and for
the same range of $x$, that is, $x=[0.02, 0.35]$. The result of
the model is,
\begin{equation}
\int^{0.345}_{0.02}dx (\bar{d}-\bar{u})=0.085,
\end{equation}
to be compared with the experimental value of $0.068\pm 0.0106$. For the
entire range in $x$, we get:  $\int^{1}_{0}dx (\bar{d}-\bar{u})=0.103$. The
corresponding experimentally extrapolated value is $0.1 \pm 0.018$, which is
in excellent agreement with our calculations. At $Q=7.35$ $GeV$, this gives
$S_{G}= 0.264$. In Figure (8), $\bar{d}(x)-\bar{u}(x)$
and $\frac{\bar{d}}{\bar{u}}$ in the proton are shown as a function of $x$ at
$Q=7.35$ $GeV$ along with the measured results. \\
\subsection{Proton Structure Function $F_{2}^{p}$}
We are now in a position to present the results for the proton structure
function, $F_{2}^{p}$. In Eqs.(17,18) we presented the form of parton
distributions in a CQ. Using those relations, along with the numerical values
given in the Appendix, the structure function of the CQ is obtained via Eq. (1).
The shape of CQ distributions in the proton is provided by Eqs. (23) and (24).
Now, all the ingredients are in place to calculate $F_{2}^{p}$ from Eq. (20).
In Figure (9) the results are shown at many values of $Q^{2}$.
As it is evident from Figure (4), sole use of CQ structure to reconstruct the
$F_{2}^{p}$ data is not enough. The CQ contribution fall a few percent short
of representing the actual data. However, as mentioned earlier, there is an
additional contribution from the {\it{inherent}} partons to $F^{p}_{2}$,
which is calculated in Eq. (31). Adding this component represents the data rather
well as can be seen from Figure (9). The data points are from \cite{20}. For
the purpose of comparing our results with other calculations, we have also
included in Figure (9), the GRV's NLO results \cite{21}, as well as the
prediction of CTEQ4M \cite{22}. Notice that we have taken the number of active
flavors to be three for $Q^{2}\leq 5 GeV^{2}$ and four, otherwise. In
Figure (10), the gluon distribution predicted by the model is presented along
with those from Ref.[16, 20].
\section{Summary and Conclusion}
In this paper we have used the notion of constituent quark as a
well defined entity being common to all hadrons. Its structure can
be calculated perturbatively to all orders in QCD. A CQ receives
its own structure by dressing a valence quark with gluon and
$q-\bar{q}$ pairs in QCD. We have calculated its structure
function in the Next-to-Leading order framework. Considering a
hadron as the bound states of these constituent quarks, we have
used the convolution theorem to extract the hadronic structure
functions for proton and pion. Besides the CQ structure
contribution to the hadrons, there is also a nonperturbative
contribution. This component is small, being only a few percent.
But it is crucial in providing a framework to understand and
explain the violation of the Gottfried sum rule and the excess of
$\bar{d}$ over $\bar{u}$ in the nucleon sea. A mechanism is
devised for this purpose and necessary calculations are performed.
We have presented the results and compared them with all the
available and relevant data,
as well as with the work of others. We found that our results are in good agreement with the experimental data. \\
{\it{Acknowledgment}} We are grateful to professor Rudolph C. Hwa for reading
the manuscript and his helpful comments and suggestions. A. N. K. is also
grateful to Semnan University partially supporting his research.
\section{APPENDIX}
In this appendix we will give the functional form of parameters of Eqs. (17, 18)
in terms of the evolution parameter, $t$. This will completely determines
partonic structure of CQ and their evolution. The results are valid for three
and four flavors, although the flavor number is not explicitly present but
they have entered in through the calculation of moments. As we explained in the
text, we have taken the number of flavors to be three for $Q^{2} \leq 5 GeV^{2}$
and four for higher $Q^{2}$ values. \\
\\
I) Valence quark in CQ (Eq. 17): \\
\\
$a= -0.1512 +1.785 t -1.145 t^{2} +0.2168 t^{3}$  \\
$b=1.460 -1.137 t +0.471 t^{2} -0.089 t^{3}$   \\
$c=-1.031 +1.037 t -0.023 t^{2} +0.0075 t^{3}$ \\
\\
II) Sea quark in CQ (Eq. 18):  \\
\\
$\alpha=0.070 - 0.213 t + 0.247 t^{2} - 0.080 t^{3}$ \\
$\beta=0.336 - 1.703 t +1.495 t^{2} - 0.455 t^{3}$ \\
$\gamma=-20.526 +57.495 t -46.892 t^{2} + 12.057 t^{3}$ \\
$\eta =3.187 - 9.141 t +10.000 t^{2} -3.306 t^{3}$ \\
$\xi=-7.914 +19.177 t - 18.023 t^{2} + 5.279 t^{3}$ \\
${\cal{N}}=1.023 +0.124 t -2.306 t^{2} +1.965 t^{3}$ \\
\\
III) Gluon in CQ (Eq. 18) \\
\\
$\alpha=0.826 - 1.643 t + 1.856 t^{2} - 0.564 t^{3}$ \\
$\beta=0.328-1.363 t + 0.950 t^{2} -0.242 t^{3}$ \\
$\gamma=-0.482 + 1.528 t -0.223 t^{2} -0.023 t^{3}$ \\
$\eta=0.480 -3.386 t + 4.616 t^{2} - 1.441 t^{3}$  \\
$\xi=-2.375 + 6.873 t -7.458 t^{2} +2.161 t^{3}$ \\
${\cal{N}}=2.247-6.903 t + 6.879 t^{2} -1.876 t^{3}$ \\
\\

\newpage
\section{Figure Caption}
Figure-1. Moments of partons in a CQ at $Q^{2}=8.5, 25, 120$ $GeV^{2}$. The dashed
curve represents the variation of leading order moments and the solid curve is
that of next to leading order.\\
Figure-2. Parton distributions in a CQ at a typical value of $Q^{2}=20$ $GeV^{2}$ \\
Figure-3. Parton distributions in proton and $\pi^{-}$ at $Q^{2}=20$ $GeV^{2}$ as
a function of $x$. Here only perturbatively generated partons are presented. \\
Figure-4 Soft gluon distribution, {\it{$xg_{soft}$}}, as a function of $x$ at various
$Q^{2}$ values. \\
Figure-5. The effect of {\it{inherent}} component to a sea parton distribution
in proton is presented as a function of $x$. The dashed-dotted line is the
perturbatively generated {\it{in-constituent quark}} contribution. The solid line
represents the sum of the two components. \\
Figure-6. Processes responsible to $SU(2)$ symmetry breaking in the nucleon sea
and violation of Gottfried sum rule. \\
Figure-7. Pion valence structure function as a function of $x$ at
$Q^{2}=5.5$ $(Gev/c)^2$. Solid curve is the result of the model
calculations
and the data points are from Ref.[17]. \\
Figure-8. The ratio $\frac{\bar{d}}{\bar{u}}$ and the difference $\bar{d}-\bar{u}$
as a function of $x$. The solid line in the model calculation and the dotted
line is the prediction of CTEQ4M [22]. Data are from Ref. [18]. \\
Figure-9. Proton structure function $F_{2}^{p}$ as a function of $x$ calculated
using the model and compared with the data from Ref. [20] for different $Q^{2}$ values.
The dot-dot-dashed line is the prediction of GRV Ref.[21] and the dashed line is
that of CTEQ4M Ref.[22]. \\
Figure-10. The gluon distribution in proton as a function of $x$ at $Q^{2}=20$ $GeV^{2}$.
We have also shown the prediction of GRV (dashed-dotted line) and CTEQ4M (dashed line).
The data points are from H1 collaboration.

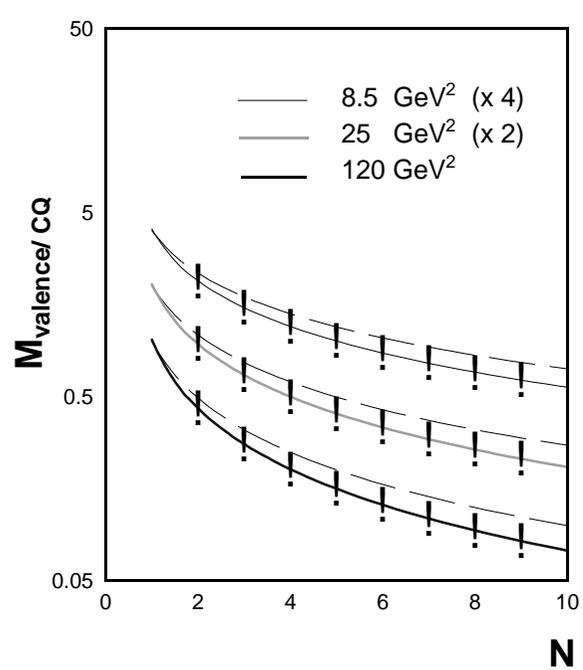
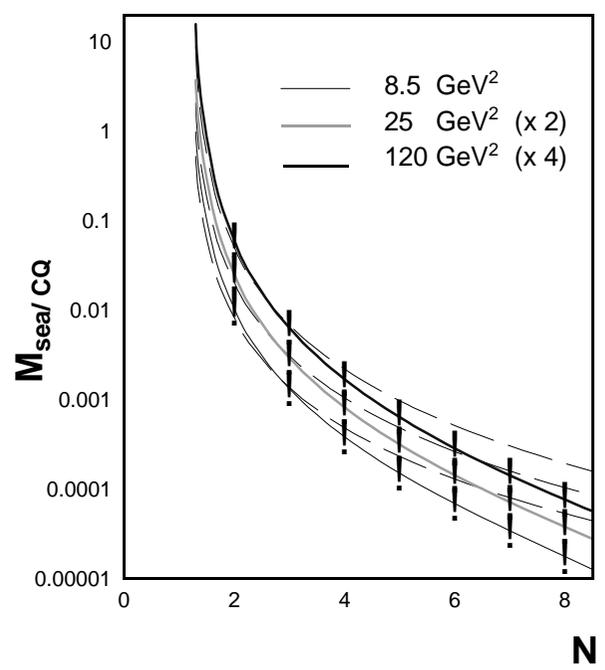
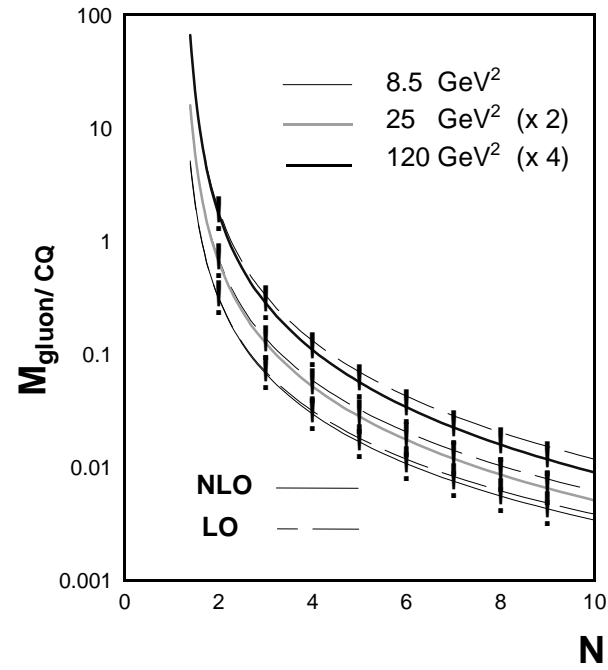
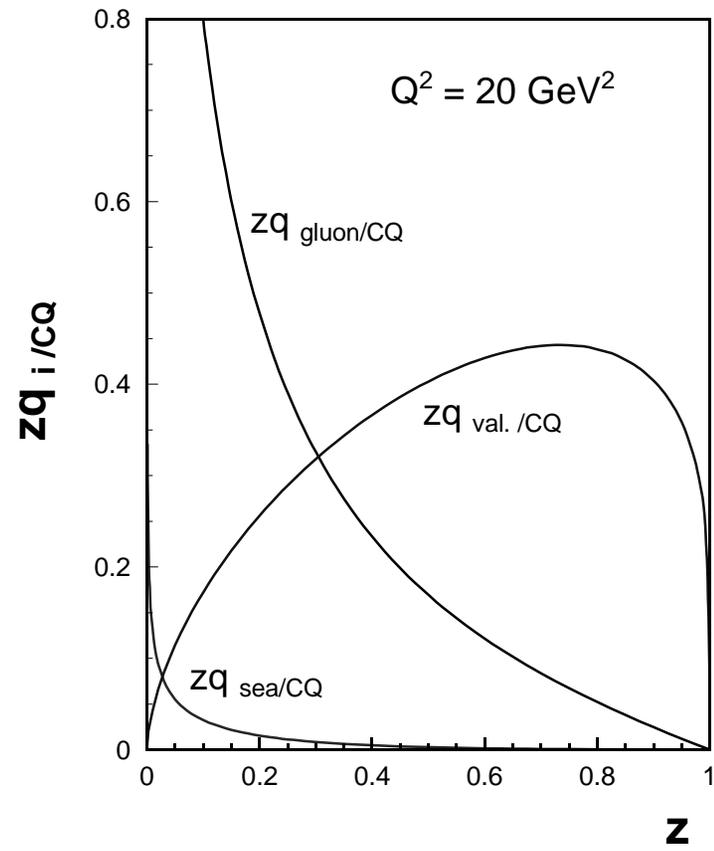
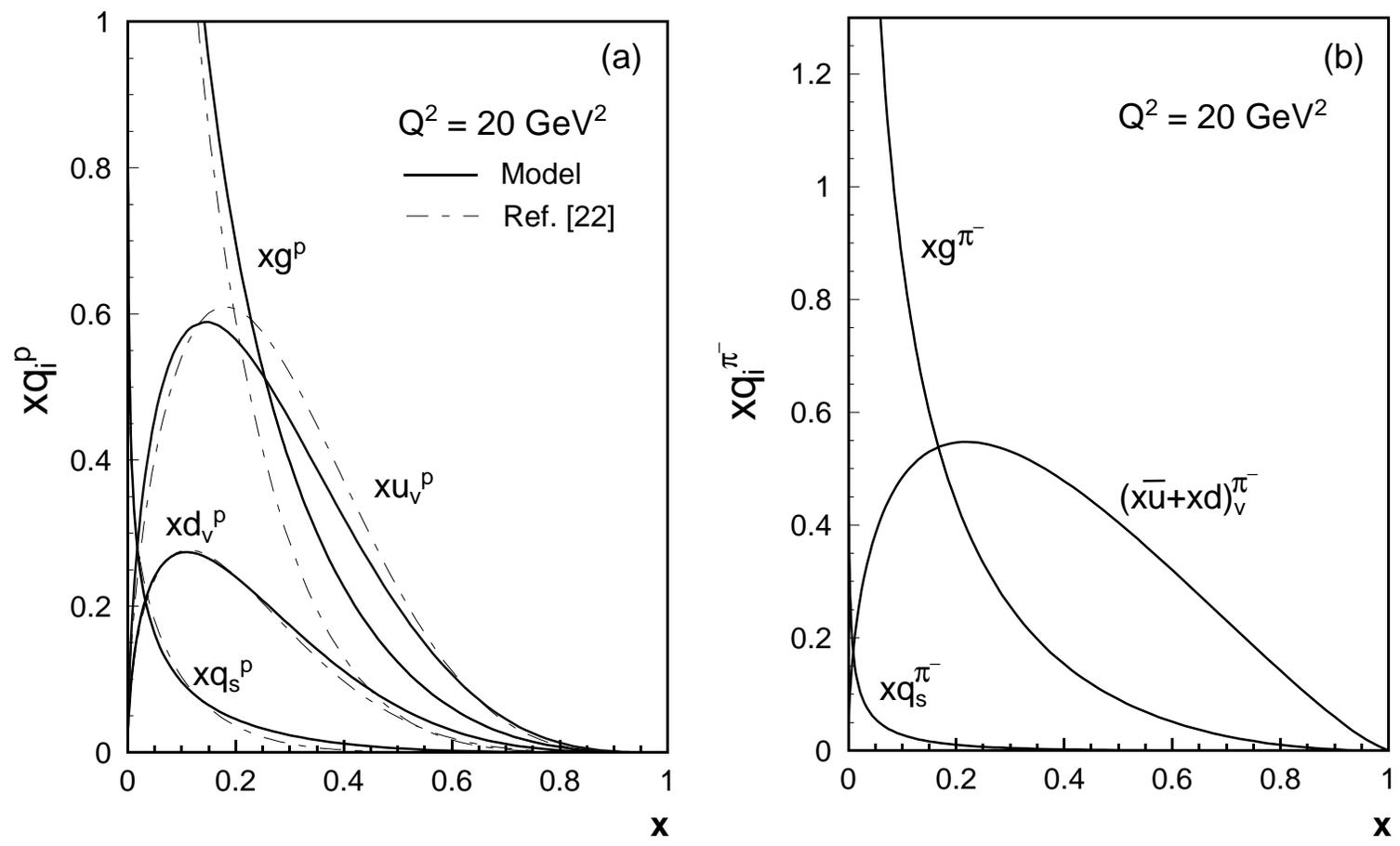
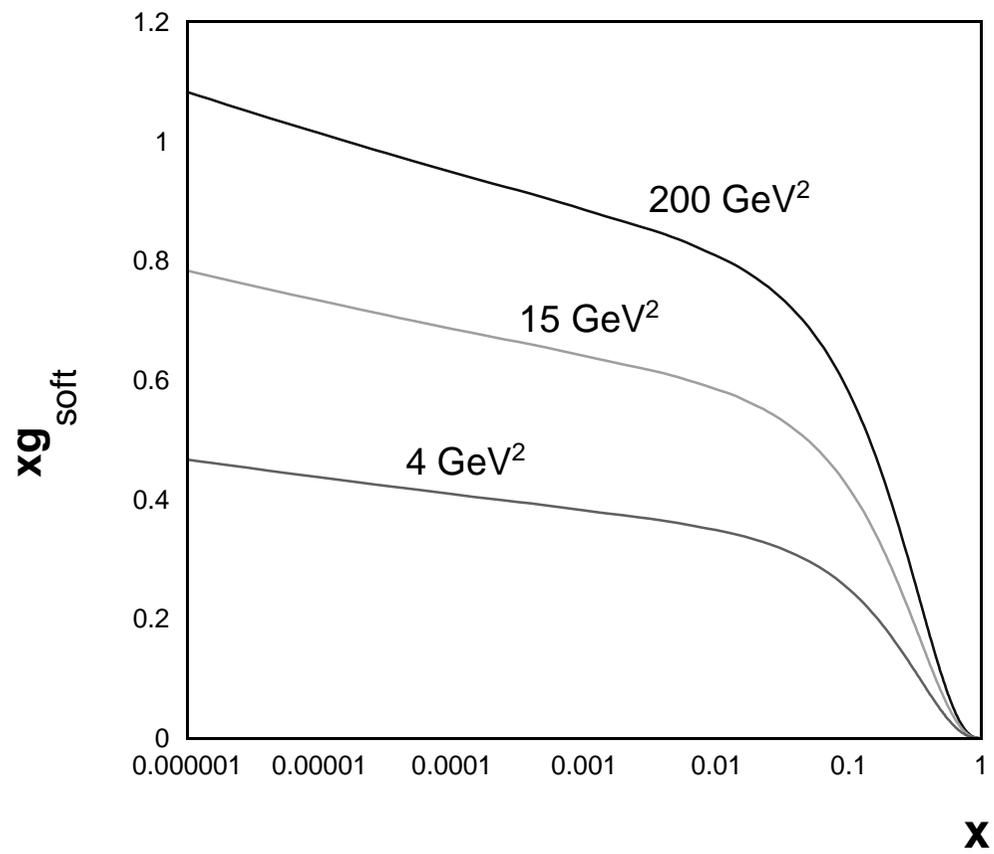
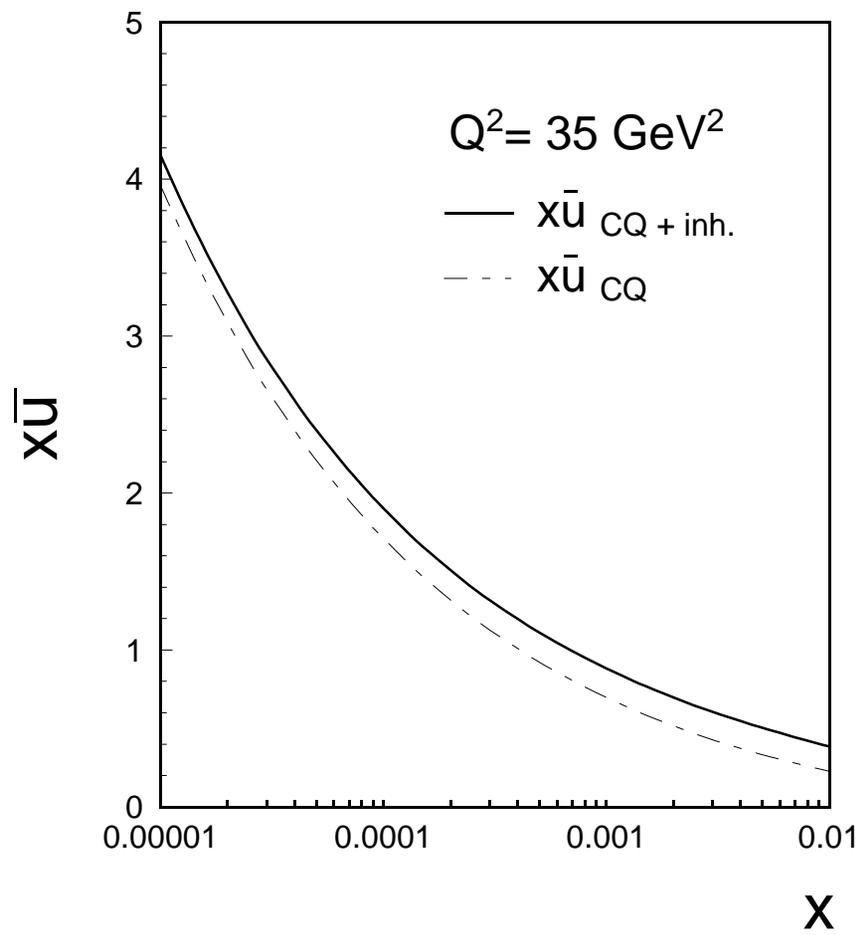
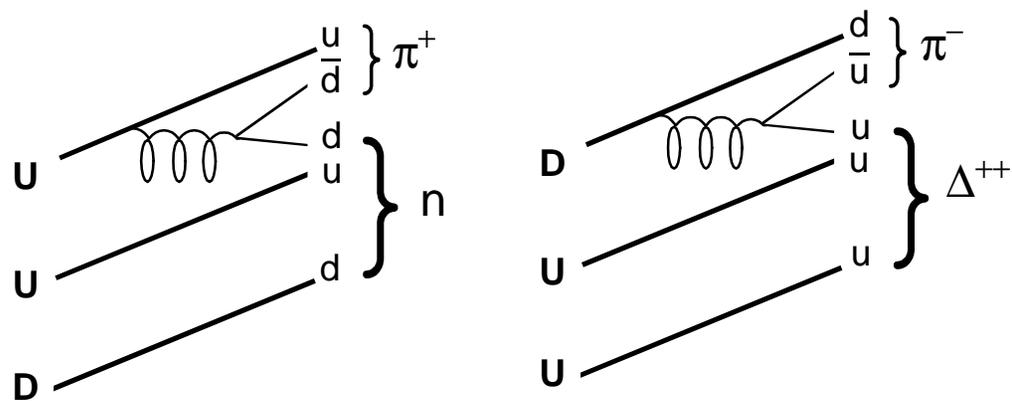
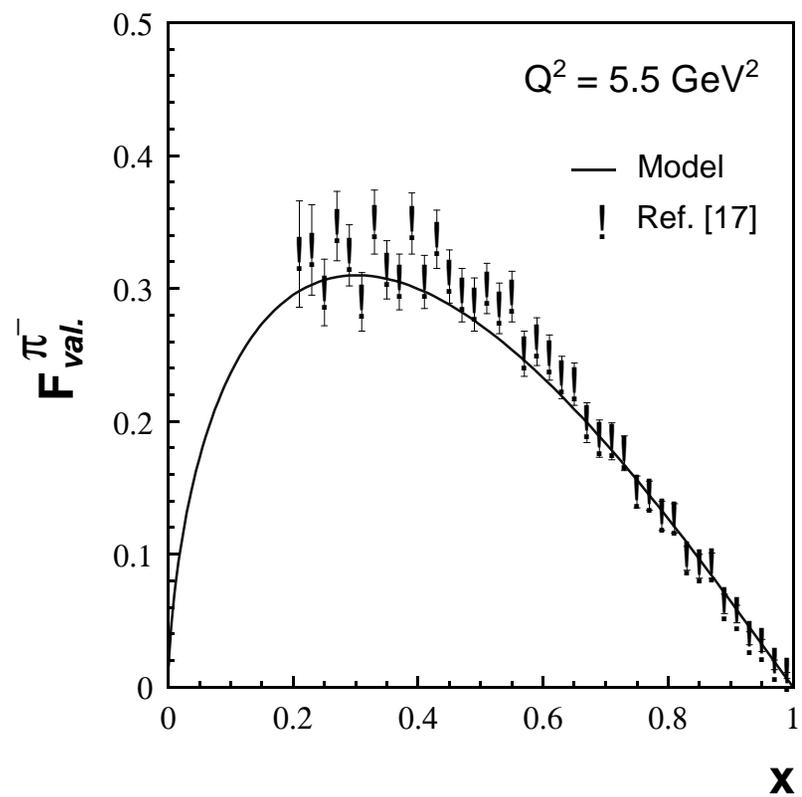
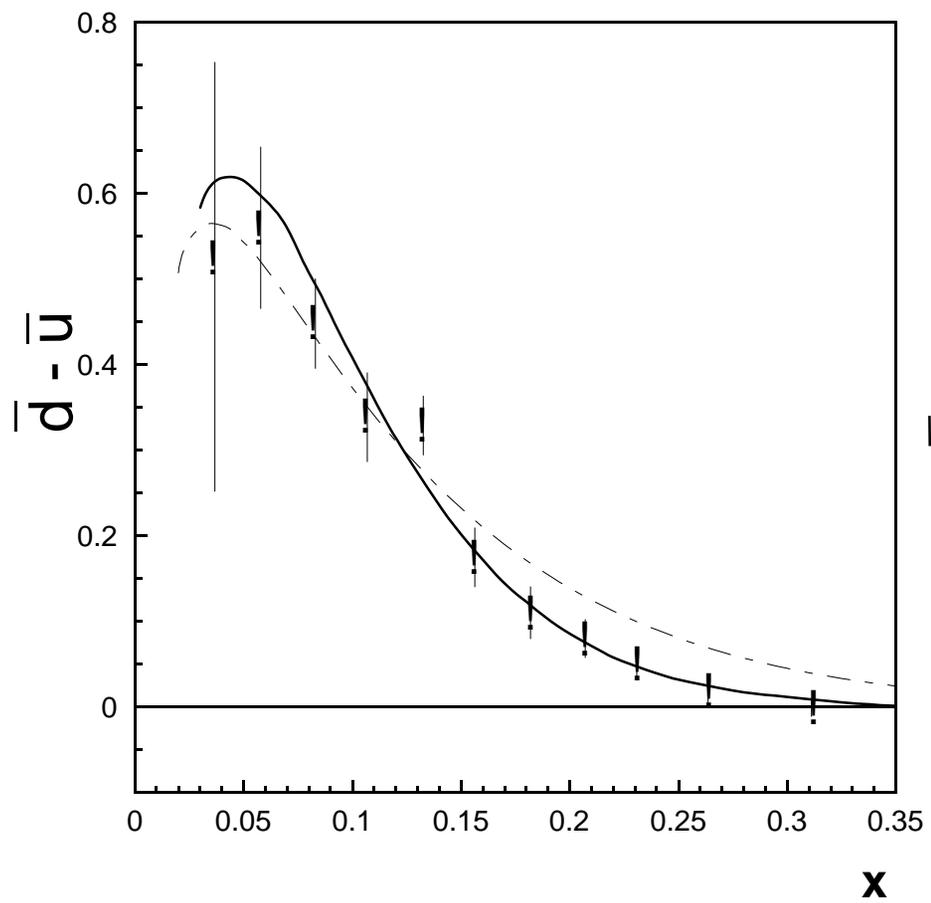
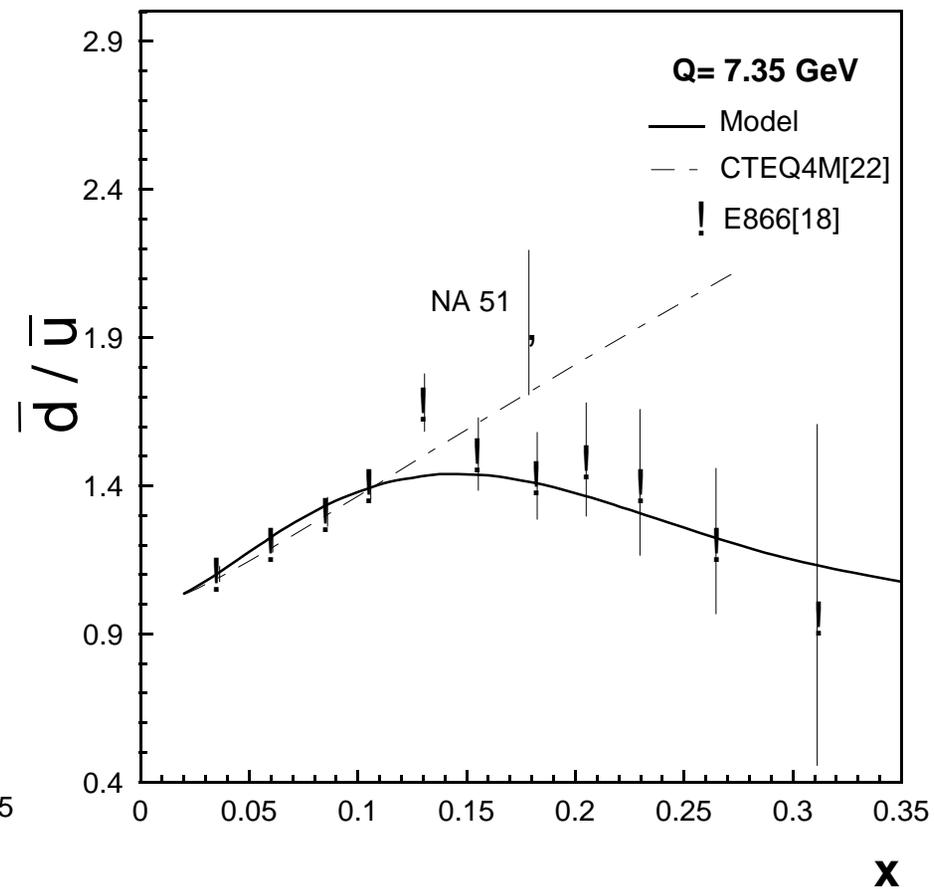
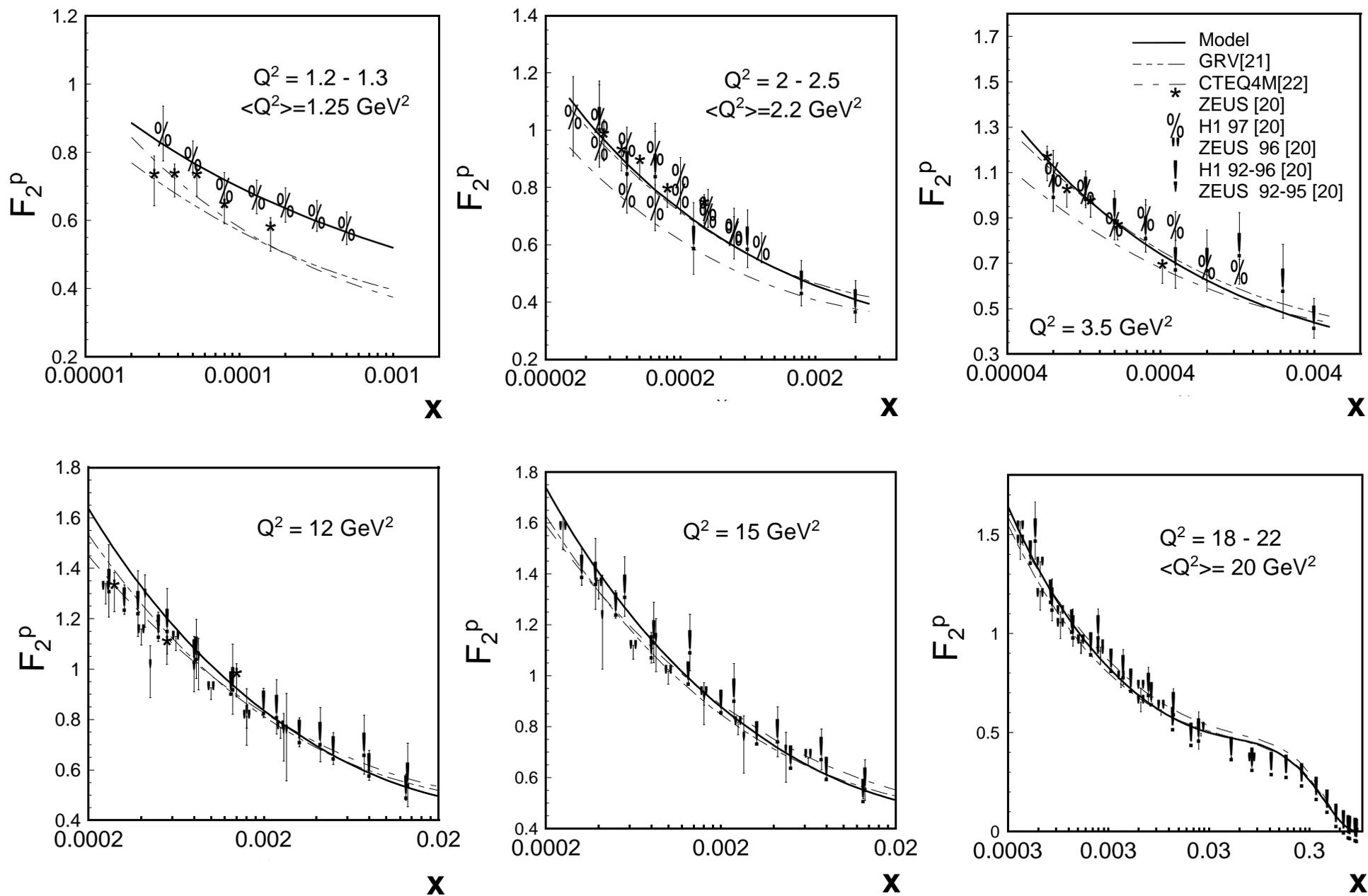
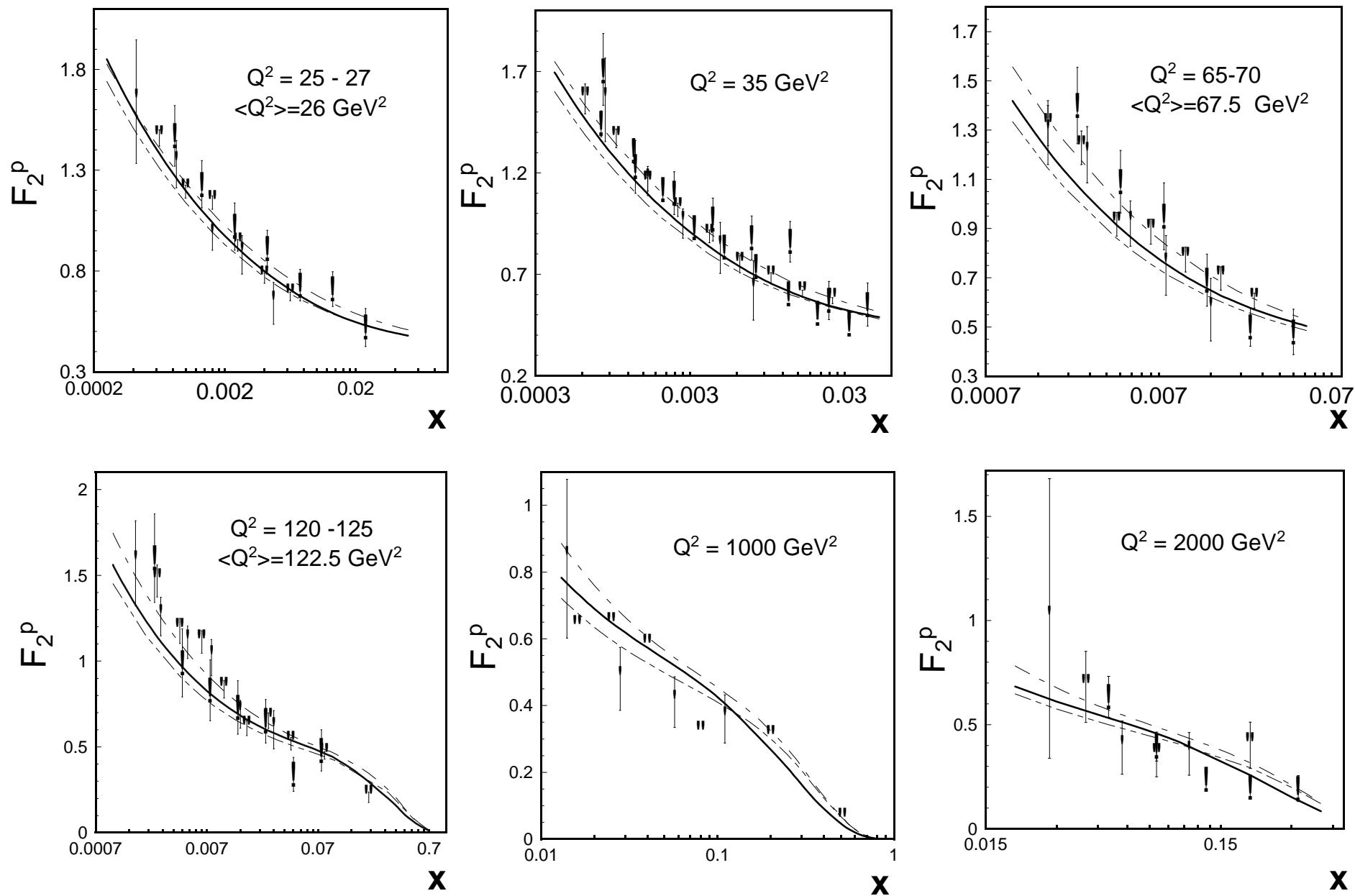
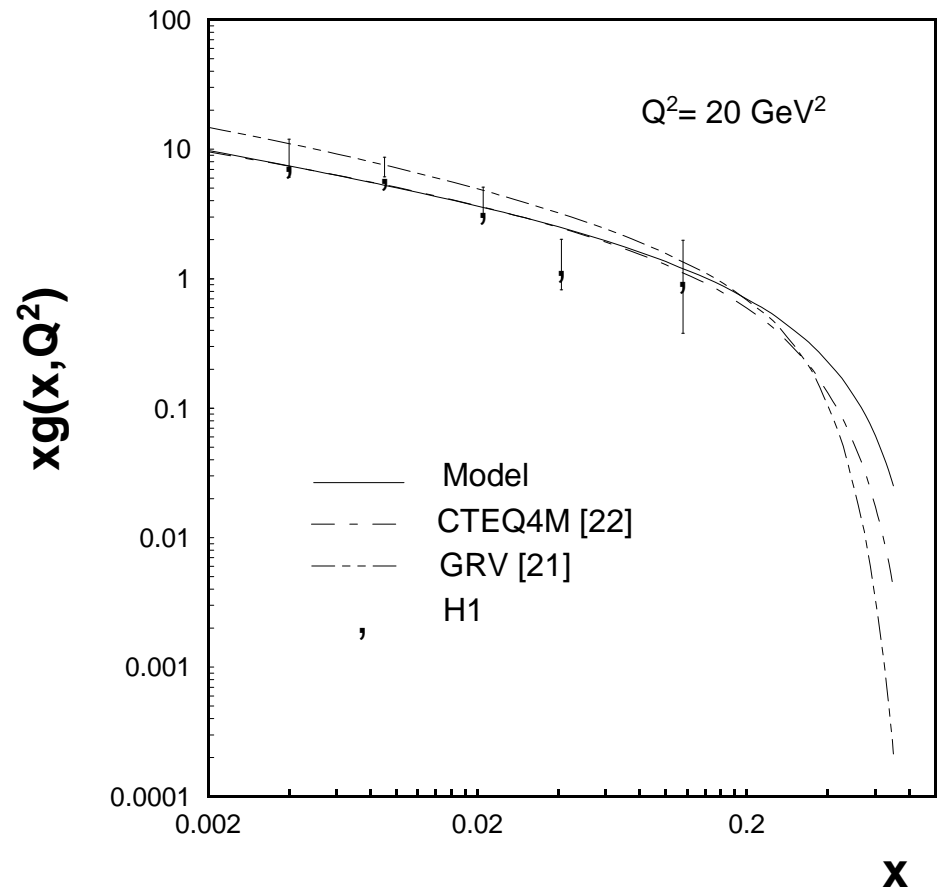
\end{document}